\documentclass[]{iopart}

\newcommand{\mathd}{\mathrm{d}}

\pdfoutput=1

\usepackage{setstack,bbm,amscd}
\usepackage{graphicx}
\begin{document}

\title{Role of correlations in the thermalization of quantum systems}

\author{A Smirne$^{1,2}$, E-M Laine$^{3}$, H-P Breuer$^{4}$, J Piilo$^{3}$ and  B Vacchini$^{1,2}$}
\address{$^{1}$Dipartimento di Fisica, Universit{\`a} degli Studi di
Milano, Via Celoria 16, I-20133 Milan, Italy}
\address{$^{2}$INFN, Sezione di Milano, Via Celoria 16, I-20133
Milan, Italy}
\address{$^{3}$Turku Centre for Quantum Physics, Department of
Physics and Astronomy, University of Turku, FI-20014 Turun
yliopisto, Finland}
\address{$^{4}$Physikalisches Institut, Universit\"at Freiburg,
Hermann-Herder-Strasse 3, D-79104 Freiburg, Germany}
\ead{andrea.smirne@unimi.it {\rm and} emelai@utu.fi}

\date{\today}

\begin{abstract}
We investigate the equilibration and thermalization properties of quantum systems interacting
with a finite dimensional environment. By exploiting the concept
of time averaged states, we introduce a completely positive map 
which allows to describe in a quantitative way
the dependence of the equilibrium state on the initial condition.
Our results show that the thermalization of quantum systems is favored
if the dynamics induces small system-environment correlations,
as well as small changes in the environment, as measured by the
trace distance.
 
 \end{abstract}

\section{Introduction}\label{sec:intro}
The mechanisms behind  thermalization have recently attracted a renewed interest and initiated the development of 
novel statistical formulations of equilibration in the realm of quantum mechanics \cite{Popescu2006, Reimann2008, Linden2009, Linden2010, Reimann2010, Lychkovskiy2010, Short2011, Gogolin2011, Riera2012}.
In all these descriptions the total system under investigation  
is associated with a finite dimensional Hilbert space and thus the asymptotic limit of the dynamics does not exist and the system returns, 
with possibly very long recurrence time, arbitrarily close to its initial state infinitely many times \cite{Bocchieri1957, Percival1961}.
Relaxation to equilibrium in the usual sense is thus impossible.
Nevertheless, one can introduce an extended notion of equilibration if the system tends towards some
state, which can be identified as equilibrium state of the dynamics, and stays close to it most of the time. There will still be some fluctuations around the equilibrium state,
but extremely small or rare.

In this work we consider the situation in which
a closed quantum system can be decomposed into two parts, an open system $S$
and a bath $B$, and investigate the equilibration
properties of the subsystem $S$. We use the extended notion of equilibration, i.e, we will say that
the open system equilibrates if its time evolved state 
(also called reduced state) $\rho_S(t)$
approaches some equilibrium state and spends
most of the time close to it.
In the same spirit, one can introduce the notion of
thermalization of the open 
system by means of additional conditions on its equilibrium
state \cite{Linden2009}. Namely, one requires that the latter does not
depend on the initial total state, beside a possible dependence 
on macroscopic parameters, such as temperature,
characterizing the initial state of the bath. In this case one says that the 
open system thermalizes if, in addition,
the equilibrium state takes the form of a Gibbs state.
The capability of an open system to thermalize
ultimately traces back to specific properties of the total Hamiltonian, which 
fixes the evolution of the total closed system and, in particular, characterizes
the interaction between the open system and the bath. 
Indeed, if the open system and the bath do not interact
no thermalization is expected. Moreover, if there
are conserved quantities for the open system its equilibrium state will
unavoidably depend on the initial reduced state.
More generally, it has been shown \cite{Gogolin2011} that the lack of a sufficient amount of entanglement in the energy eigenbasis is 
a basic reason for 
the absence of thermalization.

Here, we want to discuss the thermalization of open quantum systems
within the above-mentioned framework,
with the aim of clarifying the role of general dynamical mechanisms that can induce or prevent it.
Apart from the obvious situation of conserved quantities for the open system, which features
of the dynamics imply a dependence of the equilibrium state on the initial reduced
state? Thermalization requires a sufficient amount of entanglement in the energy eigenbasis, but what is the role
played by the interaction induced correlations between the open system and the bath? 
We investigate how the equilibrium state is modified when one resets the correlations between the system and the bath, as well as the environmental 
state, to their initial value. We find that small system-environment correlations, together with small changes in the environmental state, generally lead to
an equilibrium state that hardly depends on the initial state of the open system.
The system-environment correlations built by the dynamics thus play a role which is in a sense opposite to that
of the entanglement in the energy eigenbasis. The interaction induced correlations can prevent the thermalization: strong system-environment correlations and changes in the environmental state
allow to distinguish between equilibrium states corresponding to
different initial states. The information about the initial
  state of the reduced system, transferred to the environment through
  the establishment of correlations, does influence the equilibrium
  state. In this sense it can be considered as trapped in the
  equilibrium state.
For this reason, we will refer to this mechanism preventing thermalization as \emph{information trapping}. The quantitative characterization of information trapping will be given in terms of trace distance, which measures the
distinguishability between quantum states \cite{Fuchs1999} and has
already been used to detect through its variation the information flow
between system and environment \cite{Breuer2009,Liu2011}, as also
discussed later on.

\section{General framework}

%In this paper, we will address the issue of thermalization 
%within the context of quantum mechanics by means of the framework
%provided by the theory of open quantum systems \cite{Breuer2007}.
Consider a finite dimensional Hilbert space, which can be decomposed as
$\mathcal{H}= \mathcal{H}_S \otimes \mathcal{H}_B$, with $\mathcal{H}_S$ and $\mathcal{H}_B$
Hilbert spaces associated  with the open system $S$ and the bath $B$, respectively. 
A crucial assumption here
is that the total Hilbert space $\mathcal{H}$ has finite dimension, which will be denoted by $d$, 
while we denote by $d_S$ and $d_B$ the
dimensions of the open system and the bath. 
The evolution of the total system
is governed by a one-parameter group of unitary operators $U(t)$, which 
is fixed by the total Hamiltonian
\begin{equation}
H = H_S + H_B + H_{S B},
\end{equation}
where $H_S$ and $H_B$ are the free Hamiltonian of the system and the bath,
and $H_{SB}$ is the interaction term.
Moreover, assume a product initial total state, so that, for a fixed initial state of the bath, there is a well-defined reduced
dynamics \cite{Breuer2007}, i.e. there is a family of completely
 positive and trace preserving maps $\Lambda(t)$ on the set
 $\mathcal{S}(\mathcal{H}_S)$ of statistical operators on
 $\mathcal{H}_S$ such that the reduced state $\rho_S(t)$ at time
 $t$ is given by $\rho_S(t) = \Lambda(t) \rho_S$, where $\rho_S \equiv \rho_S(0)$:
 \begin{eqnarray}
 \Lambda(t) : \mathcal{S}(\mathcal{H}_S) &\longrightarrow&\, \mathcal{S}(\mathcal{H}_S) \nonumber\\
 \qquad \qquad  \rho_S & \longrightarrow&\, \rho_S(t) = \Lambda(t) \rho_S.
 \end{eqnarray}  
In \cite{Linden2009} it
has been shown that an open system equilibrates under very general
assumptions if the effective dimension of the bath, i.e. the dimension
of the subspace of $\mathcal{H}_B$ involved into the dynamics, is much
larger than the open-system dimension $d_S$. In this case, for any
initial state $\rho_S$, the corresponding time evolved state
$\rho_S(t)$ will be most of the time close to the time averaged state
$\overline{\rho_S}$, which is defined as
\begin{equation}
  \overline{\rho_S} =  \lim_{t \rightarrow \infty}
  \frac{1}{t} \int^t_0 \mathd \tau \rho_S (\tau)  \label{eq:ta}
\end{equation}
and represents the equilibrium state of
the reduced dynamics. Note that throughout the whole paper we will use
an overline to denote the time average of any operator or function.
More precisely, if the total Hamiltonian has nondegenerate energy gaps,
the average distance between $\rho_S(t)$ and the time averaged state $\overline{\rho_S}$
is bounded by  \cite{Linden2009}
\begin{equation} \label{eq:ww}
\overline{D(\rho_S(t),  \overline{\rho_S})} \leq \frac{1}{2}\sqrt{\frac{d_S}{d_{\mbox{eff}}(\overline{\rho_B})}},
\end{equation}
where
\begin{equation}\label{eq:deff}
d_{\mbox{eff}}(\overline{\rho_B}) = \frac{1}{\mbox{tr}_B \left\{\overline{\rho_B}^2\right\}}
\end{equation}
represents the effective dimension of the bath.
Here and in the following we characterize the distance between quantum
states by means of the trace distance. Given two quantum states $\rho^1$ and $\rho^2$, their trace distance is defined as
\begin{equation}
D(\rho^1, \rho^2)=\frac{1}{2} \| \rho^1-\rho^2 \|, 
\end{equation}
where the trace norm is considered.
The upper bound in (\ref{eq:ww}), which has been proven in \cite{Linden2009} for a pure product initial state,
can be easily extended to a mixed product initial state, see Appendix A.

On that account, the open system equilibrates under very general conditions, but nevertheless, despite some significant results \cite{Linden2009, Reimann2010, Lychkovskiy2010, Gogolin2011, Riera2012}, 
it is still a widely open question which are the conditions that determine whether an open system, besides equilibrating, thermalizes. 
In the following, we will focus on this issue and, in particular, on finding conditions on the dependence of the time averaged state
on the initial state of the open system. 
%thus showing the central role played by the correlations between the system and the bath which are due to their interaction.

\section{Information trapping}

\subsection{Time averaging map}
First of all, let us take a closer look at the time averaged state defined in (\ref{eq:ta}). In the following, we assume for simplicity a non degenerate 
Hamiltonian $H = \sum_k E_k |E_k\rangle \langle E_k|$, but analogous considerations
can be done in the degenerate case.
For a fully generic initial total state $\rho_{SB}$, one has
\begin{equation}\label{eq:ta2}
\overline{ \rho_S } = \sum_k \langle E_k| \rho_{SB} |E_k \rangle \sigma^k_S,
\end{equation} 
where the notation
\begin{equation}\label{eq:rhok}
\sigma^k_S \equiv \mbox{tr}_B\left\{|E_k\rangle \langle E_k|\right\}
\end{equation}
has been introduced.
If the initial total state is a product state $\rho_S \otimes \rho_B$, with fixed environmental state $\rho_B$,
equation (\ref{eq:ta2}) defines a map $\overline{\Lambda}$ on the state space of the open system $\mathcal{S}(\mathcal{H}_S)$:
\begin{eqnarray}\label{eq:tam}
\rho_S &\rightarrow& \overline{\Lambda} \rho_S:= \lim_{t\rightarrow \infty} \frac{1}{t} \int^t_0 \mathd \tau \rho_S(\tau) =  \sum_k \langle E_k| \rho_{S} \otimes \rho_{B} |E_k \rangle \sigma^k_S,
\end{eqnarray}
that can also be written as 
\begin{equation}\label{eq:tam2}
\overline{\Lambda} \rho_S =  \sum_k p_k \sigma^k_S,
\end{equation}
with
\begin{equation}\label{eq:pk}
p_k \equiv \langle E_k| \rho_S \otimes \rho_B | E_k \rangle = \mbox{Tr}\left\{|E_k\rangle \langle E_k|  \rho_S \otimes \rho_B\right\}.
 \end{equation}
This map associates to any initial state of the system the corresponding time averaged state,
and therefore we will call it the \emph{time averaging map}. This is a linear, trace preserving and completely positive map and its image $\overline{\Lambda}(\mathcal{S}(\mathcal{H}_S)) \equiv \mbox{Im}\, \overline{\Lambda}$ can be identified as the set of equilibrium states of the reduced dynamics.
A state $\rho_S$ will be said to be invariant if it is left unchanged by the time averaging map, i.e. if $\overline{\Lambda} \rho_S = \rho_S$.
A natural question is then whether equilibrium states are invariant, i.e.,
if, given a state $\omega_S = \overline{\Lambda} \rho_S$ for some
initial state $\rho_S$, one has $\overline{\Lambda} \omega_S = \omega_S$.
This can happen for any $\omega_S$ if and only if
the map $\overline{\Lambda}$ is a projector. That is, it satisfies the idempotence relation $\overline{\Lambda}^2 = \overline{\Lambda}$, 
where of course $\overline{\Lambda}^2$ indicates the composition of $\overline{\Lambda}$
with itself. Note that 
\begin{equation}\label{eq:osos}
\overline{\Lambda}^2 \rho_S = \sum_{k k'} \langle E_k| \rho_S \otimes \rho_B | E_k \rangle  \langle E_{k'}| \sigma^k_S \otimes \rho_B | E_{k'} \rangle \sigma^{k'}_S.
\end{equation}
%We will say that there is information trapping 
%with the idempotence property of the time averaging map is violated. Indeed, if $\overline{\Lambda}$ is not a projector
%the equilibrium state depends on the initial state of the system.
%Thus, by investigating information trapping, we get some insight into
%a general mechanism that can prevent thermalization and that, as will become more clear later, is strictly connected to the creation of
%correlations between the system and the bath due to their mutual interaction.

Let us now make the following important remark. First, recall that any trace preserving
and positive map $\Lambda$ is a contraction for the trace distance \cite{Kossakowski1972, Ruskai1994a}, i.e. $D(\Lambda \rho^1, \Lambda \rho^2) \leq D(\rho^1, \rho^2)$ for any
$\rho^1$ and $\rho^2$. A map is further said to be \emph{strictly contractive} \cite{Raginsky2002, Heinosaari2011} if
$D(\Lambda \rho^1, \Lambda \rho^2) < D(\rho^1, \rho^2)$ for any $\rho^1\neq\rho^2$. Indeed,
the time averaging map $\overline{\Lambda}$ is contractive, but in general not strictly contractive. 
It is clear that the only way for it to be both strictly contractive and idempotent is
to map every initial state to the same time averaged state. That is, the maps $\Lambda_{\rho}$ defined as
\begin{equation}
\Lambda_{\rho} \rho_S = \rho \quad \forall \rho_S \in \mathcal{S}(\mathcal{H}_S)
\end{equation}
for a fixed state $\rho$, are the only idempotent and strictly
contractive maps on $\mathcal{S}(\mathcal{H}_S)$.  In fact, let
$\omega_1, \omega_2 \in \mbox{Im} \overline{\Lambda}$ be two elements
of the image of $\overline{\Lambda}$, i.e. $\omega_1 =
\overline{\Lambda} \rho^1_S$ and $\omega_2 = \overline{\Lambda}
\rho^2_S$ for some $\rho^1_S, \rho^2_S \in
\mathcal{S}(\mathcal{H}_S)$.  The idempotence of $\overline{\Lambda}$
implies that $D(\overline{\Lambda} \omega_1, \overline{\Lambda}
\omega_2) = D(\omega_1, \omega_2)$ and, because of the strict
contractivity, it follows that $\omega_1 = \omega_2$.  Hence the image
of $\overline{\Lambda}$ consists of only a single element which proves
our claim.

As a consequence, the dependence of the equilibrium state on the
initial state of the system can be always related to the violation of either
property. In other words, the absence of thermalization can always be associated with either the lack of idempotence or strict contractivity of the time averaging map.
In the following we will focus on the violation of the idempotence of $\overline{\Lambda}$, which will be referred to 
as \emph{information trapping}. This will be shown to capture an interesting dependence of the equilibrium state
on the initial state, and we
will demonstrate a connection between information trapping and the creation of
correlations between the system and the bath due to their mutual interaction.

\subsection{Measure for information trapping}\label{sec:mit}
Rather than simply assessing whether 
the time averaging map is idempotent, one needs to quantify  its possible deviation from idempotence
in order to point out if this can be treated as "small".
The very definition of the equilibration of a quantum system interacting with a finite dimensional
bath involves the idea that the reduced state $\rho_S(t)$ will stay most of the time in
a neighborhood of the corresponding equilibrium state $\overline{\Lambda}\rho_S$.
For the sake of concreteness, let us denote as $\mathcal{X}$ the radius of such neighborhood.
Now, if the distance between two different equilibrium states $\overline{\Lambda}\rho^1_S$ and $\overline{\Lambda}\rho^2_S$
is smaller than $\mathcal{X}$, the corresponding time evolved states $\rho^1_S(t)$ and $\rho^2_S(t)$
can be close to each other for almost all times, so that one cannot practically infer
that they approach different equilibrium states by monitoring their evolution.
This leads us to the conclusion that  it is more meaningful to investigate the amount of information
trapping of a given dynamics, rather than its mere presence. 

In particular, we propose the following measure for information trapping:
\begin{eqnarray} 
\mathcal{T}(\overline{\Lambda}) &=& \max_{\substack{\rho_S \in \mathcal{S}(\mathcal{H}_S)}} D(\overline{\Lambda}^2 \rho_S , \overline{\Lambda} \rho_S).\label{eq:t1}
\end{eqnarray}
This directly quantifies the violation of the idempotence of $\overline{\Lambda}$, and it is indeed
equal to $0$ if and only if $\overline{\Lambda}$ is idempotent. In Appendix B, we introduce
an alternative, but qualitatively equivalent, measure.
Now, if $\mathcal{T}(\overline{\Lambda})$ exceeds $\mathcal{X}$,
there is some $\rho_S$ such that one can actually determine that $\rho_S$ and $\overline{\Lambda}\rho_S$ evolve to different
equilibrium states and no thermalization occurs. In addition, as will be shown by means of examples, the measure $\mathcal{T}(\overline{\Lambda})$  
provides a useful way to describe how the different features of
a given dynamics can enhance or decrease the information trapping and thus
the dependence of the equilibrium state on the initial state of the open system.

\section{Information trapping and system-environment correlations}
In this section, we explicitly connect the notion of information trapping
with the interaction induced correlations between the system and the bath, as well as the 
changes in the environmental state.
First of all, it is useful to come back to the full unitary dynamics, where the time averaging can
be described by means of a trace preserving and completely positive map $\overline{U}$, such that (compare with (\ref{eq:ta2})),
\begin{equation}\label{eq:tau}
\overline{U} \rho_{SB} = \lim_{t \rightarrow \infty}
  \frac{1}{t} \int^t_0 \mathd \tau \rho_{SB} (\tau) = \sum_k \langle E_k| \rho_{SB} |E_k \rangle |E_k\rangle \langle E_k|.
\end{equation}
Indeed, this map can be defined for any initial total state, but we will
focus on the case $\rho_{SB} = \rho_S \otimes \rho_B$, with fixed $\rho_B$, to guarantee the existence
of the reduced map $\overline{\Lambda}$, which can be expressed as
\begin{equation}
\overline{\Lambda}\rho_S = \mbox{tr}_B\left\{\overline{U} (\rho_S \otimes \rho_B)\right\}.
\end{equation}
In the following diagram
one can see the relation between the map $\overline{U}$ on the total system and both the reduced time averaging map $\overline{\Lambda}$
and its two-fold application $\overline{\Lambda}^2$:
%\begin{equation}\label{eq:dia}
%\begin{array}{ccccccc}
    % \rho_S \otimes \rho_B&\overset{\overline{U}}{\longrightarrow} &\omega_{SB} = \overline{U}( \rho_{S}\otimes \rho_B) & / &\omega_S \otimes \rho_B &\overset{\overline{U}}{\longrightarrow}&  \overline{U}( \omega_S \otimes \rho_B) \\
 %\rotatebox{270}{$\longrightarrow$}& &  \rotatebox{270}{$\longrightarrow$} & &  \rotatebox{270}{$\longrightarrow$} & &  \rotatebox{270}{$\longrightarrow$}  \\
 %\phantom{\frac{\frac{3}{3}}{3}}   \rho_S & \overset{\overline{\Lambda}}{\longrightarrow}& \omega_S = \overline{\Lambda}\rho_S & / & \omega_S = \overline{\Lambda}\rho_S & \overset{\overline{\Lambda}}{\longrightarrow}&   \overline{\Lambda}^2\rho_S,
   % \end{array}
%\end{equation}
 \begin{equation}
    \label{eq:dia}
\fl    
\begin{CD}
       \rho_S \otimes \rho_B @>\overline{U}>> \omega_{SB} = \overline{U}( \rho_{S}\otimes \rho_B) @. \qquad\,\,  \omega_S \otimes \rho_B  @>\overline{U}>> \overline{U}( \omega_{S}\otimes \rho_B)\\
       @V\mbox{tr}_B VV  @V\mbox{tr}_B VV   \qquad\,\,@V\mbox{tr}_B VV   @V\mbox{tr}_B VV\\
       \rho_S   @>\overline{\Lambda}>>      \omega_S = \overline{\Lambda}\rho_S @.  \qquad\,\,  \omega_S = \overline{\Lambda}\rho_S @>\overline{\Lambda}>>    \overline{\Lambda}^2\rho_S   \nonumber
    \end{CD}
 \end{equation}
where we introduced the notation 
$\omega_{SB} \equiv \overline{U} (\rho_{S} \otimes \rho_{B})$
to indicate the time averaged state of the total system, 
so that $\omega_S = \mbox{tr}_B {\omega_{SB}} = \overline{\rho_S}$ and $\omega_B = \mbox{tr}_S {\omega_{SB}} =  \overline{\rho_B}$ 
are the time averaged states of the system and the bath, respectively. 
In particular, note how the reduced map $\overline{\Lambda}^2$ is obtained after resetting the total state
from $\omega_{SB}$ to $\omega_S \otimes \rho_B$. Now, the map $\overline{U}$ on the total state is always idempotent, i.e., $\overline{U}^2 = \overline{U}$, as can be easily checked
by means of (\ref{eq:tau}), since it amounts to a von Neumann measurement of the total energy. Introducing the map $\overline{\Phi} =\mbox{tr}_B \circ \overline{U} : \mathcal{S}(\mathcal{H}_{SB})\rightarrow \mathcal{S}(\mathcal{H}_S)$ , 
from the diagram (\ref{eq:dia}) and the idempotence of $\overline{U}$,
one has $\overline{\Lambda}^2 \rho_S = \overline{\Phi}( \omega_S \otimes \rho_B)$, while $\overline{\Lambda} \rho_S = \overline{\Phi} (\omega_{SB})$. 
But then, since $\overline{\Phi}$ is trace preserving and completely positive , the contractivity of the trace distance implies
\begin{equation}\label{eq:bb}
D(\overline{\Lambda}^2 \rho_S, \overline{\Lambda} \rho_S) \leq D(\omega_{SB}, \omega_S \otimes \rho_B) \leq  D(\omega_{SB}, \omega_S \otimes \omega_B) + D(\rho_B, \omega_B).
\end{equation}
The information trapping is upper bounded by the total amount of correlations
between the system and the bath in the total time averaged state $\omega_{SB}$ plus the distinguishability
between the time averaged state of the bath $\omega_B$ and the fixed initial state $\rho_B$. This means that,
while a small amount of entanglement in the energy eigenbasis prevents a full thermalization \cite{Gogolin2011},
such phenomenon will be generally favored if the dynamics builds up
a small amount of correlations between the system and the bath, together with
small changes in the state of the bath.
 
From a physical point of view, we can explain the connection 
between information trapping and system-environment correlations by taking advantage of the notion of information
flow associated with the changes of the trace distance between reduced
states in the course of time \cite{Breuer2009, Liu2011}.
The basic idea is that if there is some information trapped
into the open system when it approaches the equilibrium, then, by resetting 
the system-environment correlations as well as the bath state to their
initial condition, see (\ref{eq:dia}), 
one can restart an information flow between the system and the bath, thus leading the system
to a different equilibrium state. The distinguishability between the new equilibrium state $\overline{\Lambda}^2 \rho_S$
and $\overline{\Lambda} \rho_S$ then provides a way to quantify the information trapped into the open
system due to the system-environment correlations and the changes in the environmental state.

% Note the analogy between
% the relation in Eq.(\ref{eq:bb}) and the upper bound introduced in \cite{Laine2010} for the increase
% of the trace distance $D(\rho^1_S(t), \rho^2_S(t))$ above its initial value due to system-environment correlations and different
% environmental states in the initial total states $\rho^1_{SE}$ and $\rho^2_{SE}$. Nevertheless, we emphasize
% that the inequality in Eq.(\ref{eq:bb}) is referred to different quantities and has a very
% different physical meaning with respect to the upper bound in \cite{Laine2010}. In fact, it concerns the effects of the overall 
% system-environment correlations and changes in the bath
% due to the full dynamics on the equilibrium state of the open system.
The relevance of bounds, determined by correlations in the total state
as well as different environmental states, for the trace distance
among different system states has been first pointed out in
\cite{Laine2010}, where the time dependence of the trace distance has
been related to the presence of initial correlations. Here however the
different system states do not correspond to different initial
conditions, but rather to the action of distinct mappings.

\section{Examples}

\subsection{Product energy eigenbasis} \label{sec:pei}
As a first representative example, consider a product energy eigenbasis \cite{Linden2009}, 
\begin{equation}\label{eq:hk1k2}
H = \sum_{k_1 k_2} E_{k_1 k_2} |E_{k_1}\rangle \langle E_{k_1}| \otimes  |E_{k_2}\rangle  \langle E_{k_2}|.
\end{equation}  
For such an Hamiltonian any reduced observable
of the form $A = \sum_{k_1} a_{k_1} |E_{k_1}\rangle \langle E_{k_1}|$ represents a conserved quantity. 
Note that a non degenerate conserved quantity on the open
system implies a product eigenbasis of the total Hamiltonian.
For $H$ as in (\ref{eq:hk1k2}), the time averaging map is not strictly contractive.
In fact, one has
\begin{equation}\label{eq:k1k2}
 \omega_S = \overline{\Lambda} \rho_S = \sum_{k_1} \langle E_{k_1}| \rho_S|E_{k_1}\rangle |E_{k_1}\rangle \langle E_{k_1}|,
\end{equation}
implying that if we choose as initial states two different elements of the basis $\left\{|E_{k_1}\rangle\right\}_{k_1 = 1, \ldots d_S}$,
$\rho^1_S = |E_{j_1}\rangle \langle E_{j_1}|$ and $\rho^2_S = |E_{l_1}\rangle\langle E_{l_1}|$, we get
$$D(\overline{\Lambda} |E_{j_1}\rangle \langle E_{j_1}|, \overline{\Lambda} |E_{l_1}\rangle \langle E_{l_1}|) = D(|E_{j_1}\rangle \langle E_{j_1}|, |E_{l_1}\rangle \langle E_{l_1}|) =1.$$
On the other hand, for a product energy eigenbasis there is no
information trapping, since the time averaging map is a projection.
Even more, as it clearly appears from the expression of the time averaged state (\ref{eq:k1k2}), by setting $\omega_S \otimes \rho_B$
as initial total state, the reduced system does not evolve at all. 
%Explicitly,
%\begin{eqnarray}
%\nonumber
%&&\mbox{tr}_B \left\{U(t) (\omega_S \otimes \rho_B) U^{\dag}(t)\right\} =
% \sum_{k_1}  \langle E_{k_1}| \rho_S|E_{k_1} \rangle  |E_{k_1}\rangle \langle E_{k_1}| = \omega_S. \nonumber
%\end{eqnarray}
This clearly shows that, unlike violation of strict contractivity, information trapping describes a mechanism
preventing thermalization which is not merely due to conserved quantities of the open system.

Moreover, for 
a product energy eigenbasis, the total time averaged state
is a product state, i.e. $\omega_{SB} = \omega_S \otimes \omega_B$, but in
general the time averaged state of the bath $\omega_B$ will be different from the initial state $\rho_B$:
\begin{equation} \label{eq:bbb}
D(\omega_B, \rho_B) = \frac{1}{2}\| \sum_{k_2 \neq k_2'} \langle E_{k_2}| \rho_B| E_{k'_2}\rangle |E_{k_2}\rangle \langle E_{k'_2}| \|.
\end{equation}
Nevertheless, we have just shown that $\overline{\Lambda}^2 \rho_S = \overline{\Lambda} \rho_S $ for any $\rho_S$, whichever
the value $D(\omega_B, \rho_B)$ in equation (\ref{eq:bbb}).
Indeed, inequality (\ref{eq:bb}) gives an upper bound to the amount of
information trapping which implies that
it may well happen that, despite strong system-environment correlations or changes in the environmental state,
the equilibrium state presents no information trapping. 

\subsection{Jaynes-Cummings model}
Let us now consider the Jaynes-Cummings model, i.e. a two-level system
interacting under the rotating wave approximation
with a single mode of the radiation field. Moreover, the latter is
initially in a thermal state, so that
the effective dimension of the bath can be made arbitrarily large
by properly increasing the bath temperature.
Indeed, this model is much 
simpler than systems
with a macroscopic number of degrees of freedom \cite{Reimann2008, Reimann2010, Short2011} or many-body systems \cite{ Gogolin2011,Rigol2008, Cramer2008},
which are usually taken into account when studying thermalization in the
quantum setting. In this context, the 
Jaynes-Cummings model can be seen as a toy model, which allows us
to explicitly evaluate all the quantities
presented in the previous sections. We emphasize, however, that our general analysis can
%is by no means restricted to such simple situations.
be applied to any open system, the only requirements being that
the dimension of the total Hilbert space is finite,
and that the open system and the bath are initially uncorrelated.

The Hamiltonian giving the total dynamics is
\begin{equation}\label{eq:hjc}
H =  \omega_0 \sigma_+ \sigma_- +  \omega b^{\dag} b+g \left( \sigma_+ \otimes b + \sigma_- \otimes b^{\dag} \right),
\end{equation}
where $\sigma_+ = |1 \rangle \langle 0|$
and $\sigma_- = |0 \rangle \langle 1|$ are the raising and lowering operators of
the two-level system, while the creation and annihilation operators of the field mode, $b^{\dag}$ and $b$, obey the
standard bosonic commutation relation. Finally, $g$ is the coupling constant
and we will denote by $\Delta = \omega_0 - \omega$ the detuning
between the frequency $\omega_0$ of the atom and the frequency $\omega$ of the field mode.
Moreover, one can think of an high-energy cutoff in order to keep the dimension of the bath finite.
For an initial total state $\rho_{SB} = \rho_S \otimes \rho_B$, where 
\begin{equation}
\rho_S = \left(\begin{array}{cc}
      \rho_{11}& \rho_{10} \\
    \rho_{01} & \rho_{00}     
    \end{array} \right) 
\end{equation}    
and $\rho_B = e^{- \beta \omega b^{\dag} b} / Z$ is the thermal state of the bath,
the reduced state at time $t$ is given by \cite{Smirne2010}
\begin{equation}
%\fl 
\rho_S(t) =  \left(\begin{array}{cc}
     \rho_{00} \left( 1 -
  \alpha (t) \right) + \rho_{11}\beta \left( t
  \right) & \rho_{10} \gamma \left( t
  \right) \\
    \rho_{01} \gamma^* \left( t
  \right) & \rho_{00} \alpha \left( t
  \right) + \rho_{11}\left( 1 - \beta (t)
  \right)
    \end{array} \right),\label{eq:rhot}
\end{equation}
where
\begin{eqnarray*}  
\alpha (t)  &=  \langle c^{\dag} \left( \hat{n}, t \right) c
  \left( \hat{n}, t \right) \rangle_B
   \\
  \beta (t)  &=  \langle c^{\dag} \left( \hat{n} + 1, t \right)
  c \left( \hat{n} + 1, t \right) \rangle_B
 \\
  \gamma (t)  &=  \langle \mathbbm{} c \left( \hat{n}, t \right)
  c_{} \left( \hat{n} + 1, t \right) \rangle_B,
 \end{eqnarray*}
%  $\alpha (t)  =  \langle c^{\dag} \left( \hat{n}, t \right) c
%  \left( \hat{n}, t \right) \rangle_B, 
%  \beta (t)  =  \langle c^{\dag} \left( \hat{n} + 1, t \right)
%  c \left( \hat{n} + 1, t \right) \rangle_B$ and $\gamma (t)  =  \langle \mathbbm{} c \left( \hat{n}, t \right)
%  c_{} \left( \hat{n} + 1, t \right) \rangle_B$,
  with $\langle A \rangle_B = \mbox{Tr}\left\{A \rho_B\right\}$, the
  number operator $\hat{n} = b^{\dag} b$ and 
  $$
  c \left( \hat{n}, t \right) =  e^{- i \omega t / 2} \left[\cos \left(
  \sqrt{\Delta^2 + 4 g^2 \hat{n}} \frac{t}{2} \right)-  \frac{i \Delta}{\sqrt{\Delta^2 + 4 g^2 \hat{n}}}\sin \left( \sqrt{\Delta^2 + 4 g^2 \hat{n}}
  \frac{t}{2} \right)   \right] .
  $$
The time average 
%$ \lim_{t\rightarrow \infty} \frac{1}{t} \int^t_0 \mathd \tau \rho_S(\tau)$
can be directly calculated, thus giving
\begin{eqnarray}  
\overline{\Lambda} \rho_S &=& \left(\begin{array}{cc}
     \rho_{00} \left( 1 -
  \overline{\alpha} \right) + \rho_{11}\overline{\beta} &0 \\
    0 & \rho_{00} \overline{\alpha}  + \rho_{11}\left( 1 - \overline{\beta}
  \right)
    \end{array} \right),\label{eq:tajc}
\end{eqnarray}
with
\begin{eqnarray}
\nonumber  \overline{\alpha} = \left\langle \frac{\Delta^2 + 2 g^2 \hat{n}}{\Delta^2+4 g^2 \hat{n}} \right\rangle_B\label{eq:ab}\\
\overline{\beta} = \left\langle \frac{\Delta^2 + 2 g^2 (\hat{n}+1)}{\Delta^2+4 g^2 (\hat{n}+1)}\right\rangle_B.\label{eq:ab2}
\end{eqnarray}
From (\ref{eq:tajc}) one has
\begin{eqnarray}
\overline{\Lambda}^2 \rho_S &=&  \left(\begin{array}{cc}
(\overline{\Lambda}^2 \rho_S)_{11} &0 \\
    0 &1-  (\overline{\Lambda}^2 \rho_S)_{11} 
\end{array}\right),    
    \nonumber\\
(\overline{\Lambda}^2 \rho_S)_{11}  &=&  \rho_{00}  \left( 1 -
  \overline{\alpha} \right)(\overline{\alpha} +\overline{\beta}) + \rho_{11}\left(1+(\overline{\alpha}+\overline{\beta})(\overline{\beta}-1) \right)
\end{eqnarray}
so that $\overline{\Lambda}^2 = \overline{\Lambda}$ if and only if $\overline{\alpha}+\overline{\beta} = 1$ or
$\overline{\alpha}=1$ and $\overline{\beta} = 1$. The latter case corresponds
to $g=0$, which implies, as expected from the discussion in section (\ref{sec:pei}), 
that there is no strict contractivity, see equation (\ref{eq:rhot}), and $\overline{\Lambda}$ is idempotent.
In all the other situations one has $\overline{\beta}+\overline{\alpha}-1<1$ and the map is strictly contractive.
Moreover, since for $\Delta \neq 0$, $\frac{\Delta^2 + 2 g^2 \hat{n}}{\Delta^2+4 g^2 \hat{n}} + \frac{\Delta^2 + 2 g^2 (\hat{n}+1)}{\Delta^2+4 g^2 (\hat{n}+1)} > 1$,
the only possibility to have $\overline{\alpha}+\overline{\beta} = 1$ is actually the resonant situation, $\Delta = 0$.
In this case the derivation of $\overline{\alpha}$ through equation (\ref{eq:ab}) has to be performed
quite carefully. One can take the limit $\Delta \rightarrow 0$ into the function the series in (\ref{eq:ab})
converges to, or equivalently note that $\hat{c}^{\dag}(0,t)\hat{c}(0,t) = 1$,
so that for $\Delta=0$ one has
\begin{equation}\label{eq:alim}
\overline{\alpha} = \frac{1}{Z} + \sum_{n>0} \frac{1}{2} \frac{e^{-\beta \hbar \omega n }}{Z} = \frac{1}{2 Z} + \frac{1}{2}.
\end{equation}
Thus, we have $\overline{\alpha} + \overline{\beta} - 1 = 1 / (2Z)$, meaning that, apart from the trivial case, in
this model there is always information trapping and there never is only one equilibrium state.

\begin{figure}[h]
\begin{center}
\includegraphics[scale=0.7]{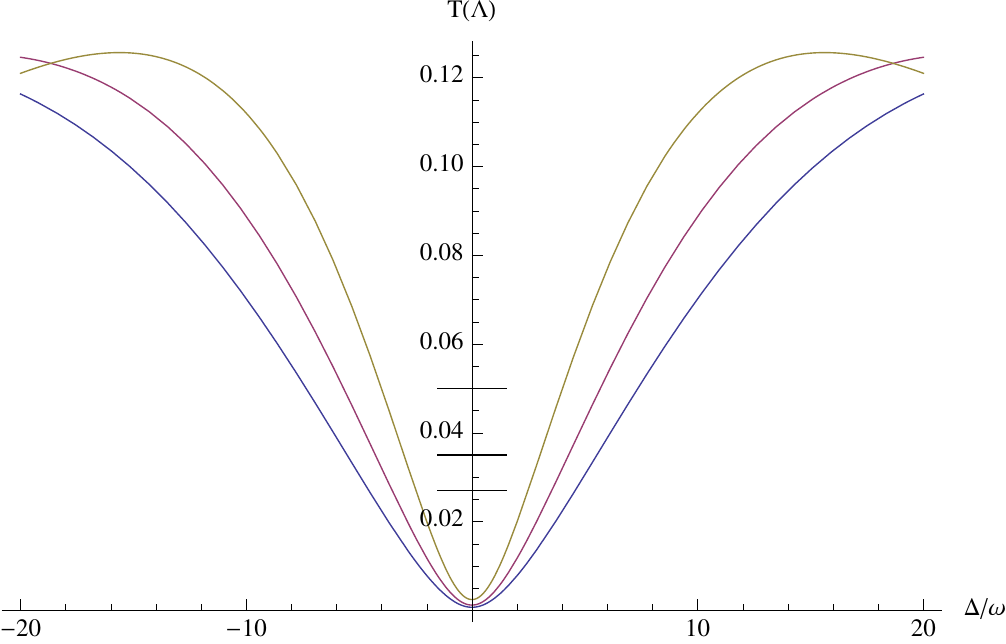}
\caption{{\small{Measure for information trapping $\mathcal{T}(\overline{\Lambda})$ 
defined in (\ref{eq:ex1}) as a function of $\Delta/\omega$ for $g/\omega = 1$ and $\beta \omega = 0.003$ (blue line),
$\beta \omega =0.005$ (red line) and $\beta \omega =0.01$ (yellow line);
the values for $\Delta = 0$ can be obtained through equation (\ref{eq:tlim}), as well. 
The marks on the vertical axis give the values of the r.h.s. of (\ref{eq:ww}), which upper bounds
the value of $\mathcal{X}$ as discussed in the text, for $\Delta=0$
and for the different temperatures: these marks are, respectively, 
$0.027, 0.035$ and $0.050$.}}}
\label{fig:1}
\end{center}
\end{figure}

We can now characterize the dependence of the equilibrium state of the open system
on its initial state by means of the measure for information trapping introduced in equation (\ref{eq:t1}) of section (\ref{sec:mit}):
\begin{eqnarray}
\nonumber \mathcal{T}(\overline{\Lambda}) &=& \max_{\rho_{11}}|\rho_{11}(\overline{\alpha}+\overline{\beta} - 2)(\overline{\alpha}+\overline{\beta} - 1)+(1-\overline{\alpha})(\overline{\alpha}+\overline{\beta} - 1)|\\
&=& \left(\overline{\alpha}+\overline{\beta} - 1\right) (1-\overline{\beta}),\label{eq:ex1}
\end{eqnarray}
where the maximum is assumed for $\rho_S = |1\rangle \langle 1|$.  Note that from (\ref{eq:alim}) one has for $\Delta = 0$
\begin{equation}\label{eq:tlim}
\mathcal{T}(\overline{\Lambda}) = \frac{1}{4}(1-e^{-\beta \omega}).
\end{equation}
In figure (\ref{fig:1}), 
we have plotted the measure $\mathcal{T}(\overline{\Lambda})$ as a function of the detuning $\Delta$,
for different values of the bath temperature $T$. The resonant situation $\Delta = 0$
represents a minimum 
%for the amount of information trapping and then, since $\overline{\Lambda}$ is strictly contractive,
for the dependence of the equilibrium state on the initial state. 
The residual amount of information trapping has to be compared with the radius of equilibration $\mathcal{X}$, 
as discussed in section (\ref{sec:mit}).
If $\mathcal{T}(\overline{\Lambda})<\mathcal{X}$,
the residual dependence of the
equilibrium state on the initial state of the open system is not enough 
to recognize that $\rho_S$ and $\overline{\Lambda}\rho_S$ evolve to different equilibrium states.
On the ground of numerical simulations, we can consider
the right hand side of (\ref{eq:ww}) as an upper bound to $\mathcal{X}$, see also \cite{Gogolin2011}.  
Thus, in figure (\ref{fig:1}) one can see that the information trapping is actually larger than $\mathcal{X}$ for high enough detuning.

From the point of view of the Hamiltonian eigenvectors the condition $\Delta = 0$ is in fact very peculiar: the eigenvectors of the 
Jaynes-Cummings Hamiltonian, the so-called dressed states,
reduce for $\Delta = 0$ to
\begin{equation}
|\Psi_n^{\pm}\rangle = \frac{1}{\sqrt{2}}\left(|0, n\rangle \pm |1, n-1\rangle \right),
\end{equation}
plus the vacuum state $|0,0\rangle$. Every eigenvector $|\Psi_n^{\pm}\rangle$ is maximally entangled in $\mathbbm{C}^2 \otimes \mathbbm{C}^2_n$,
where $\mathbbm{C}^2_n$ is the two-dimensional subspace of
$\mathcal{H}_B$ spanned by 
$|n\rangle$ and $|n-1\rangle$.
Note that, at resonance, both the entanglement on the energy eigenbasis and the
amount of residual information trapping, see (\ref{eq:tlim}), do not depend on the coupling constant $g$ between the system
and the bath. 
Thus, for the model at hand, a high entanglement in the energy eigenbasis ensures an (effective) independence of
the equilibrium state on the initial state of the open system, as
one can expect from \cite{Gogolin2011}. 

\begin{figure}[h]
\begin{center}
\includegraphics[scale=0.7]{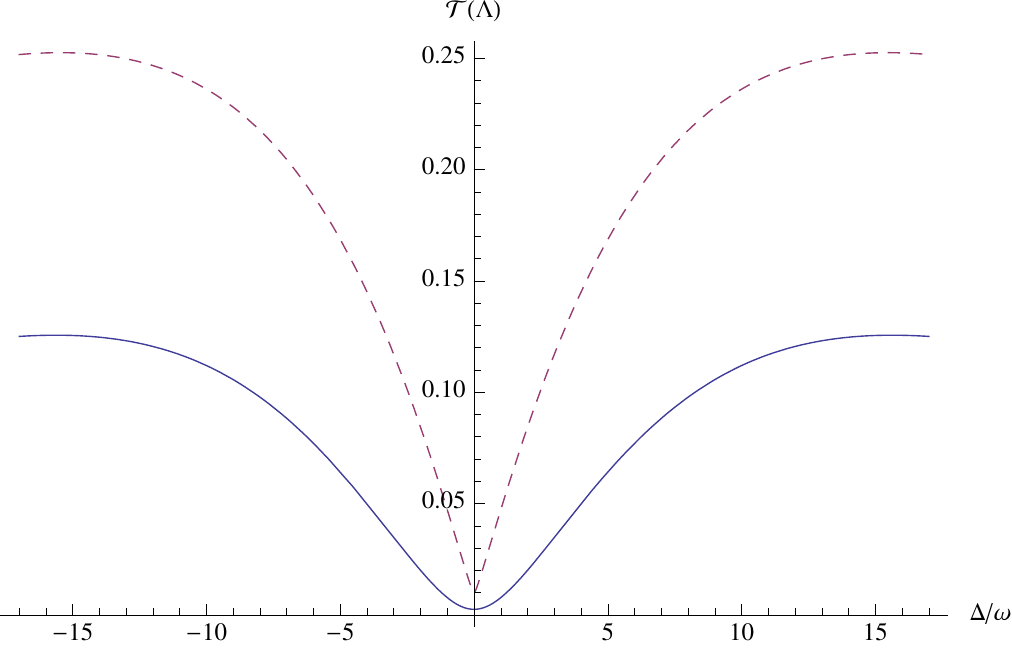}
\caption{{\small{Measure for information trapping $\mathcal{T}(\overline{\Lambda})$ (blue line)
and $D(\omega_{SB}, \omega_S \otimes \omega_B)+D(\omega_B, \rho_B)$ (red, dashed line)  vs. $\Delta/\omega$ for $g/\omega = 1$ and $\beta \omega = 0.01$;
$\omega_{SB}$ is given by $\overline{U}( \rho_{S}\otimes \rho_B)$, see (\ref{eq:dia}), with  $\rho_B = e^{- \beta \omega \hat{n}} / Z$ and $\rho_S=|1\rangle \langle 1|$
the reduced state maximizing the information trapping in (\ref{eq:ex1}).
}}}
\label{fig:2}
\end{center}
\end{figure}

Now, we want to explicitly quantify the role of the system-environment correlations, 
as well as the changes in the environmental
state, by means of the upper bound introduced in (\ref{eq:bb}), i.e.
$D(\omega_{SB}, \omega_S \otimes \omega_B)+D(\omega_B, \rho_B)$.
This quantity can be explicitly evaluated by following the same strategy employed in \cite{Smirne2010b} 
to calculate the amount of correlations in the total Gibbs state, which takes advantage
of the block diagonal structure of the total Hamiltonian (\ref{eq:hjc})
with respect to the dressed states. In figure (\ref{fig:2})
one can see the measure $\mathcal{T}(\overline{\Lambda})$ as a function of the detuning $\Delta$ compared
with the upper bound.
We observe how the latter, despite being quite far from the actual
value of the measure for information trapping, follows its behavior from a qualitative point of view. 
Indeed, as follows from the bound (\ref{eq:bb}), small system-environment
correlations and changes in the state of the bath
imply a small amount of information trapping, and therefore
an equilibrium state of the two-level system that hardly depends
on its initial state.
But for the model
at hand, whenever strict contractivity holds, we have in addition
that the more the interaction induces system-environment
correlations and changes in the environmental state, the more the
equilibrium state will depend on the initial reduced state $\rho_S$. 
%This is indeed a stronger result than the general upper bound in (\ref{eq:bb}). 

\subsection{Structured environment}\label{sec:struct}
 
As a complementary example, we consider now the model 
of a system interacting with a structured reservoir
introduced in \cite{Gemmer2004, Gemmer2006}. A two-level system is coupled
to two energy bands with the same width $\delta \epsilon$; the energy levels in each band are equidistant 
and there are $N_1$ ($N_2$) levels in the lower (upper) band. The distance $\Delta E$ between the central levels of the two bands is in resonance
with the free energy of the two-level system. The coupling constants between the two-level system and the two bands
are independent and identically distributed
complex Gaussian random variables and their overall strength is parametrized by a constant $\lambda$.
Finally, we assume that the initial state of the environment is given by a maximally mixed combination of the lower band levels. 
Thus, by means of Hilbert space averaging \cite{Gemmer2006} or correlated projection superoperators \cite{Breuer2006}
techniques, one gets the following equations in the Schr{\"o}dinger picture for the excited state
population $\rho_{11}(t)$ and the coherence $\rho_{10}(t)$:
\begin{eqnarray}
\dot{\rho}_{11}(t) &=& -\gamma \rho_{11}(t)+\gamma_1 \rho_{11}(0) \nonumber\\
\dot{\rho}_{10}(t) &=& -(i \Delta E + \gamma_2/2)\rho_{10}(t),
\end{eqnarray}
where $\gamma_i = 2 \pi \lambda^2 N_i / \delta\epsilon$, $i=1,2$, and $\gamma= \gamma_1+\gamma_2$.
These equations are solved by
%\begin{eqnarray}\label{eq:hh}
%\fl \rho_S = \left(\begin{array}{cc}
%    \rho_{11}&\rho_{10} \\
%    \rho_{01} &\rho_{00}
%    \end{array} \right) \rightarrow && \rho_S (t) = \Lambda(t) \rho_S \\ &&= \left(\begin{array}{cc}
%    \frac{\rho_{11}}{\gamma}(\gamma_1+\gamma_2 e^{-\gamma t})&\rho_{10}e^{-(\gamma_2/2+ i \Delta E) t} \\
%    \rho_{01}e^{-(\gamma_2/2-i \Delta E) t} &\rho_{00}-  \rho_{11}(\frac{\gamma_1}{\gamma}+\frac{\gamma_2}{\gamma}e^{-2 \gamma t}-1)
%    \end{array} \right). \nonumber
%\end{eqnarray}
\begin{equation}\label{eq:hh}
\rho_S (t) =  \left(\begin{array}{cc}
    \rho_{11}\frac{1}{\gamma}(\gamma_1+\gamma_2 e^{-\gamma t})&\rho_{10}e^{-(\gamma_2/2+ i \Delta E) t} \\
    \rho_{01}e^{-(\gamma_2/2-i \Delta E) t} &\rho_{00}-  \rho_{11}(\frac{\gamma_1}{\gamma}+\frac{\gamma_2}{\gamma}e^{-2 \gamma t}-1)
    \end{array} \right). \nonumber
\end{equation}
% Note that this provides an approximated description of the dynamics of the two-level system \cite{Gemmer2006, Breuer2006};
% still, we can of course introduce the notion of information trapping
% and the measure to quantify it.
The time averaging map corresponding to the evolution in (\ref{eq:hh}) is, see (\ref{eq:tam}),
\begin{equation}\label{eq:tag}
\overline{\Lambda} \rho_S = \left(\begin{array}{cc}
    \rho_{11} \frac{\gamma_1}{\gamma}&0 \\
    0 &\rho_{00} + \rho_{11}(1-\frac{\gamma_1}{\gamma})
    \end{array} \right).
\end{equation}
Indeed, for $\gamma_2 / \gamma_1 = 0$ this map reduces to the identity map, while in all the other situations it is a strictly contractive non-idempotent map.
The square of the time averaging map is given by
\begin{equation}
\overline{\Lambda}^2 \rho_S = \left(\begin{array}{cc}
      \rho_{11} \frac{\gamma^2_1}{\gamma^2}&0 \\
    0 &\rho_{00} + \rho_{11}(1-\frac{\gamma^2_1}{\gamma^2})
    \end{array} \right),
\end{equation}
so that the measure for information trapping defined in (\ref{eq:t1}) is
\begin{eqnarray}\label{eq:gemmer3} 
\mathcal{T}(\overline{\Lambda})= \frac{\gamma_1}{\gamma}-\frac{\gamma^2_1}{\gamma^2} = \frac{N_1 N_2}{(N_1+N_2)^2} 
\end{eqnarray}
and the maximization is obtained with $\rho_S = |1\rangle \langle 1|$. 
The information trapping is completely determined by the ratio $N_1/N_2$ and, in particular, it vanishes only in the limit
$N_2/N_1 \rightarrow 0$, which corresponds to the trivial situation $\overline{\Lambda}=\mathbbm{1}$, or
in the limit $N_1/N_2\rightarrow 0$.
% Note that this is the case also if $N_1, N_2 \rightarrow \infty$,
% that is also if the levels of the two energy bands 
% have a continuum distribution, at fixed width $\delta \epsilon$ of the two
% bands. 
% Thus, also
% in this continuum limit the information trapping can be non-zero
% thus providing a dynamical reason for the absence of a full thermalization.

Finally, let us present a remark about the connection
between information  trapping and the non-Markovianity
of a quantum dynamics \cite{Breuer2009, Rivas2010}. Note that
the relation between the asymptotic state of a reduced dynamics and its non-Markovianity
has been studied in \cite{Chruscinski2010}.
From (\ref{eq:hh}) one can easily obtain
a time-local master equation in the form
\begin{equation}
 \frac{\mathd}{\mathd t} \rho(t) = K(t) \rho(t)
\end{equation}
to characterize the dynamics of the two-level system. 
The time-local generator $K(t)$ is in fact simply given by \cite{Smirne2010} 
\begin{equation}\label{eq:k}
K(t) = \dot{\Lambda}(t) \Lambda^{-1}(t),
\end{equation}
so that, for the model at hand it reads
\begin{eqnarray}\label{eq:gemmer}
\fl \frac{\mathd}{\mathd t} \rho(t) &=& - i \Delta E \left[ \sigma_+ \sigma_-, \rho(t)\right] \\
\fl&&+\Gamma_1(t)\left[\sigma_- \rho(t) \sigma_+ - \frac{1}{2}\left\{\sigma_+ \sigma_- , \rho(t)\right\}\right]+\Gamma_2(t)\left[\sigma_z \rho(t) \sigma_z - \rho(t) \right],\nonumber
\end{eqnarray}
with
\begin{equation}\label{eq:gemmer2}
\Gamma_1(t) = \frac{\gamma_2 \gamma}{\gamma_1 e^{\gamma t}+\gamma_2} \qquad \qquad
\Gamma_2(t) = \frac{\gamma_1 \gamma_2}{4}\left(\frac{1-e^{-\gamma t}}{\gamma_2 e^{-\gamma t}+\gamma_1}\right).
\end{equation}
%\begin{eqnarray}\label{eq:gemmer2}
%\nonumber \Gamma_1(t) = \frac{\gamma_2 \gamma}{\gamma_1 e^{\gamma t}+\gamma_2} \\
%\Gamma_2(t) = \frac{\gamma_1 \gamma_2}{4}\left(\frac{1-e^{-\gamma t}}{\gamma_2 e^{-\gamma t}+\gamma_1}\right).
%\end{eqnarray}
Such coefficients are positive at every time, implying that the 
reduced dynamics under consideration is always Markovian,
both in the sense that it implies a monotonic decrease of the trace distance in the course of time
and in the sense that it is fixed by a divisible family of dynamical maps  \cite{Breuer2009, Rivas2010, Laine2010b,Vacchini2011a}.
This clearly shows that  one can actually
have information trapping also in the presence of a Markovian dynamics: more generally,
the dependence of the equilibrium state of the open system on its initial state 
does not provide a signature of non-Markovianity according to the above mentioned definitions.
 
 \section{Conclusions}

 In this paper, we have investigated the thermalization of finite
 dimensional quantum systems, within the framework of the theory of
 open quantum systems.  By only assuming an initial product state, we
 have shown how one can introduce a time averaging map on the state
 space of the open system that associates to any initial state the
 corresponding equilibrium state. In this way, we could formulate
 relevant questions related to equilibrium properties of the open
 system in terms of suitable properties of the time averaging map. In
 particular, the dependence of the equilibrium state on the initial
 reduced state can be always traced back to the violation of at least
 one of the properties of strict contractivity and idempotence of the
 time averaging map. Indeed the violation of idempotence has been
 shown to provide an indication on the amount of information about the
 initial system state stored in the equilibrium state.  We have
 therefore dubbed this violation as information trapping. It has been
 shown to be strictly connected to the interaction induced
 correlations between the system and the bath, as well as the changes
 in the environmental state, which keep track of system-environment information flow.  More precisely, small
 system-environment interactions, together with small changes in the
 state of the bath, lead to an equilibrium state with a small
 dependence on the initial state of the open system, as quantified by
 means of the trace distance.

Furthermore, we have introduced a measure in order to evaluate 
the amount of information trapping of a given dynamics. 
This provides a way to determine how the different
features of the dynamics influence the dependence of the equilibrium
state on the initial state of the open system and therefore how they can favor
or prevent a full thermalization. 
In particular, in the Jaynes-Cummings model
one can conclude that if the time averaging map is strictly contractive, 
then strong system-environment correlations
and changes in the environmental state imply a significant dependence
of the equilibrium state on the initial state of the open system. Indeed, it would 
be important to determine whether, or at least to what extent,
this implication holds in general.

Finally, let us note that the present results could provide
a further insight into the role of the weak coupling assumption
into the process of thermalization. If the open
system and the bath are weakly coupled, one expects that
the total state at a generic time can be effectively described
by  neglecting the system-environment correlations
and the changes in the environmental state. In this regard, it will
be of interest to investigate the connection between the correlation
properties of the total time averaged stated studied in
this work and the correlation properties of the total state
in the course of time.

\ack 
%\section*{Acknowledgments}
EML acknowledges financial support from Graduate School of Modern
Optics and Photonics and JP from Jenny and Antti Wihuri Foundation,
Magnus Ehrnrooth Foundation and Academy of Finland (mobility from
Finland 259827). AS and BV acknowledge financial support from MIUR
under PRIN 2008, and HPB from the German Academic Exchange Service
(DAAD). AS thanks for the hospitality Kimmo Luoma and everybody at the
Turku Centre for Quantum Physics, where part of this work was done
with grant COST-STSM-ECOST-STSM-MP1006-130212-012161 of the COST
Action MP1006. JP and BV
also acknowledge financial support from the COST Action MP1006.\\

\appendix

%\section{Proof of (\ref{eq:ww}) for a generic initial product state}
\section{Extension of the bound in the average distance between a state and its time average to a generic initial product state}
Here we prove the inequality
(\ref{eq:ww}) for a generic initial product state $\rho_S \otimes \rho_B$,
under the assumption that the total Hamiltonian $H$ has non degenerate energy gaps, i.e.
$E_k - E_{k'} = E_{j}-E_{j'}$ implies $k = k'$ and $j = j'$ or $k = j$ and $k' = j'$.
We will essentially follow the
proof in \cite{Linden2009} for initial pure product states.
First, let us introduce the notation
\begin{equation}
c_{k k'} = \langle E_k| \rho_S \otimes \rho_B | E_{k'}\rangle,
\end{equation}
so that
\begin{equation}\label{eq:app21}
\rho_S(t) = \sum_{k k'} e^{-i(E_k-E_{k'})t} c_{k k'} \mbox{tr}_B\left\{|E_k\rangle \langle E_{k'}|\right\}
\end{equation}
and therefore, compare with Eqs.(\ref{eq:ta2}) and (\ref{eq:rhok}),
\begin{equation}\label{eq:app22}
\overline{\Lambda}\rho_S = \sum_{k} c_{k k} \mbox{tr}_B\left\{|E_k\rangle \langle E_{k}|\right\}.
\end{equation} 
From the bound $\| \rho \|\leq \sqrt{d_S}\| \rho \|_{HS}$, where $\| \cdot \|_{HS}$ denotes the Hilbert-Schmidt norm $\| A \|^2_{HS}=\mbox{tr}_S A^\dagger A$,
and
%on the trace distance $D(\rho_S^1, \rho_S^2) \leq \sqrt{d_S \mbox{tr}_S\left\{(\rho^1_S - \rho^2_S)^2\right\}}/2$ and
by exploiting the concavity of the square root
%, one has \cite{Linden2009}
we have
\begin{equation}\label{eq:app23}
\overline{ D(\rho_S(t), \overline{\Lambda} \rho_S) } \leq \frac{1}{2}\sqrt{d_S \overline{\mbox{tr}_S\left\{(\rho_S(t) - \overline{\Lambda}\rho_S)^2\right\}}}.
\end{equation}
From Eqs.(\ref{eq:app21}) and (\ref{eq:app22}), it follows that
$$
\rho_S(t)- \overline{\Lambda}\rho_S = \sum_{k \neq k'} e^{-i(E_k-E_{k'}) t} c_{k k'} \mbox{tr}_B\left\{|E_k\rangle \langle E_{k'}|\right\}.
$$
Using the identity
%Moreover, since $H$ has non degenerate energy gaps and \cite{Linden2009}
$$
\mbox{tr}_S\left\{\mbox{tr}_B\left\{|E_{k}\rangle \langle E_{k'}|\right\}\mbox{tr}_B\left\{|E_{k'}\rangle \langle E_{k}|\right\}\right\} = \mbox{tr}_B\left\{\mbox{tr}_S\left\{|E_{k}\rangle \langle E_{k}|\right\}\mbox{tr}_S\left\{|E_{k'}\rangle \langle E_{k'}|\right\}\right\}, 
$$
and the fact that  $H$ has non degenerate energy gaps
one finds
\begin{eqnarray}
\nonumber
\fl \overline{\mbox{tr}_S\left\{(\rho_S(t)-\overline{\Lambda} \rho_S)^2\right\} }&=&
\sum_{k\neq k'} c_{k k'}c_{k' k} \mbox{tr}_B\left\{\mbox{tr}_S\left\{|E_k\rangle \langle E_k| \right\} \mbox{tr}_S\left\{|E_{k'}\rangle \langle E_{k'}| \right\} \right\}\\
%\fl \nonumber &=& \sum_{k k'} c_{k k'}c_{k' k} \mbox{tr}_B\left\{\mbox{tr}_S\left\{|E_k\rangle \langle E_k| \right\} \mbox{tr}_S\left\{|E_{k'}\rangle \langle E_{k'}| \right\} \right\} \nonumber\\
%\fl &&- \sum_kc^2_{k k'} \mbox{tr}_B\left\{(\mbox{tr}_S\left\{|E_k\rangle \langle E_k| \right\})^2  \right\} \nonumber\\
\fl &\leq& \sum_{k k'} c_{k k'}c_{k' k} \mbox{tr}_B\left\{\mbox{tr}_S\left\{|E_k\rangle \langle E_k| \right\} \mbox{tr}_S\left\{|E_{k'}\rangle \langle E_{k'}| \right\}\right\}. \nonumber
\end{eqnarray}
%Now, the trace of the product of two statistical operators is
%positive; note in particular, as a side remark, that
%\begin{equation}\label{eq:p2}
% \mbox{tr}_B\left\{\rho^k_B \rho^{k'}_B \right\} = \langle E_k | \mathbbm{1} \otimes \rho^{k'}_B |E_k \rangle.
%\end{equation}
%Then, since
Further exploiting the
Schwarz inequality
\begin{equation}\label{eq:p1}
c_{k k'}c_{k' k} \leq c_{k k} c_{k' k'}
\end{equation}
we finally come to 
\begin{eqnarray}
\fl \overline{ \mbox{tr}_S\left\{(\rho_S(t)-\overline{\Lambda} \rho_S)^2\right\}}  &\leq&  \sum_{k k'} c_{k k}c_{k' k'} \mbox{tr}_B\left\{\mbox{tr}_S\left\{|E_k\rangle \langle E_k| \right\} \mbox{tr}_S\left\{|E_{k'}\rangle \langle E_{k'}| \right\}\right\} \nonumber\\
\fl &=&\mbox{tr}_B\left\{\sum_{k} c_{k k} \mbox{tr}_S\left\{|E_k\rangle \langle E_k| \right\} \sum_{k'} c_{k' k'}\mbox{tr}_S\left\{|E_{k'}\rangle \langle E_{k'}| \right\} \right\} \nonumber\\
\fl &=&\mbox{tr}_B\left\{\overline{ \rho_B(t) }^2\right\}, 
\end{eqnarray}
which, together with (\ref{eq:app23}) and (\ref{eq:deff}), gives (\ref{eq:ww}).

\section{An alternative measure for information trapping}
The measure for information trapping in equation (\ref{eq:t1})
directly quantifies the effect of removing correlations and resetting the environmental state
to its initial condition in the time averaged state $\omega_{SB}$, see also (\ref{eq:dia}) and (\ref{eq:bb}).
On the other hand, it is in some sense arbitrary to consider only the two-fold application of
the time averaging map instead of a high number of applications.
It is in fact clear that the dynamics obtained by resetting the total state to $\overline{\Lambda} \rho_S \otimes \rho_B$
can still present some information trapped into the new equilibrium state as a consequence of further system-environment
correlations built up by the interaction, and so on. For this reason, we introduce the following alternative
measure for information trapping:
% \begin{eqnarray} 
% \mathcal{T}_{\infty}(\overline{\Lambda}) &=& \max_{\substack{\rho_S \in \mathcal{S}(\mathcal{H})}} D( \lim_{k\rightarrow \infty}\overline{\Lambda}^k \rho_S , \overline{\Lambda} \rho_S),\label{eq:t2}
% \end{eqnarray}
\begin{eqnarray} 
\mathcal{T}_{\infty}(\overline{\Lambda}) &=& \max_{\substack{\rho_S \in \mathcal{S}(\mathcal{H}_S)}} D( \lim_{k\rightarrow \infty}\overline{\Lambda}^k \rho_S , \overline{\Lambda} \rho_S),\label{eq:t2}
\end{eqnarray}
which is set equal to $1$ if the limit does not exist for some
$\rho_S$.  It is important to note that the two measures,
$\mathcal{T}(\overline{\Lambda})$ and
$\mathcal{T}_{\infty}(\overline{\Lambda})$, give the same qualitative
characterization of information trapping, i.e. also
$\mathcal{T}_{\infty}(\overline{\Lambda})$ is equal to $0$ if and only
if $\overline{\Lambda}$ is idempotent.  The "if" part is
obvious. To check the "only if" part, $\mathcal{T}_{\infty}(\overline{\Lambda})=0$
implies by definition the existence of the
limit in (\ref{eq:t2}), and therefore in particular
 \begin{equation}\label{eq:lim}
 \lim_{k\rightarrow \infty} \|\overline{\Lambda}^k \rho_S - \overline{\Lambda}^{k-1} \rho_S \| = 0 \qquad \forall\,\rho_S\in \mathcal{S}(\mathcal{H}_S).
 \end{equation}
Moreover, we can define the map $\overline{\Lambda}^{\infty}$ through
$\overline{\Lambda}^{\infty}\rho_S=\lim_{k\rightarrow
  \infty}\overline{\Lambda}^k \rho_S$,
and $\mathcal{T}_{\infty}(\overline{\Lambda}) = 0$ further
implies that $\overline{\Lambda}^{\infty}$ is equal to $\overline{\Lambda}$.
This requires $\mbox{Im} \overline{\Lambda}^{\infty} = \mbox{Im} \overline{\Lambda}$, but then,
since in general $\mbox{Im} \overline{\Lambda} \subseteq \mbox{Im}
\overline{\Lambda}^{2}\subseteq  \ldots \subseteq \mbox{Im} \overline{\Lambda}^{k}$,
it follows that $\mbox{Im} \overline{\Lambda} = \mbox{Im} \overline{\Lambda}^{2}= \ldots= \mbox{Im} \overline{\Lambda}^{k} $. 
Finally, (\ref{eq:lim}) is equivalent to $\lim_{k\rightarrow\infty} \overline{\Lambda}\arrowvert_{\mbox{Im} \overline{\Lambda}^{k-1}} = \mathbbm{1}$
implying $ \overline{\Lambda}\arrowvert_{\mbox{Im} \overline{\Lambda}} = \mathbbm{1}$, i.e. $\overline{\Lambda}^2 = \overline{\Lambda}$, which completes our proof.

Furthermore, note that
if $\overline{\Lambda}$ is strictly contractive, due to the Banach fixed point theorem it has a unique invariant
state $\rho^0_S =  \lim_{k\rightarrow \infty}\overline{\Lambda}^k \rho_S$, which is then
a natural reference state to quantify the dependence of the equilibrium state on the initial condition: in this case,
the measure (\ref{eq:t2}) can be simply written as 
\begin{eqnarray} 
\mathcal{T}_{\infty}(\overline{\Lambda}) &=& \max_{\substack{\rho_S \in \mathcal{S}(\mathcal{H})}} D( \rho^0_S , \overline{\Lambda} \rho_S).\label{eq:t22}
\end{eqnarray}
For example, in the Jaynes-Cummings model, for $g \neq 0$, the time averaging map $\overline{\Lambda}$ (\ref{eq:tajc}) is strictly contractive. 
Since
\begin{equation}
\left(\overline{\Lambda}^r \rho_S\right)_{11} =  \
    \left( 1 -
  \overline{\alpha} \right)\sum^{r-1}_{l=0}(\overline{\beta}+\overline{\alpha}-1)^l +\rho_{11}(\overline{\beta}+\overline{\alpha}-1)^r, 
\end{equation}
the limit map $\overline{\Lambda}^{\infty}$ is given by 
\begin{equation}
\overline{\Lambda}^{\infty} \rho_S =  \left(\begin{array}{cc}
   \frac{ 1 -
  \overline{\alpha}}{2-\overline{\beta}-\overline{\alpha}} &0\\
    0 &\frac{1 -\overline{\beta}}{2-\overline{\beta}-\overline{\alpha}} 
    \end{array} \right),
\end{equation}
which then provides the unique invariant state of the strictly contractive map $\overline{\Lambda}$.
The measure (\ref{eq:t2}) is thus given by
\begin{eqnarray}
\mathcal{T}_{\infty}(\overline{\Lambda}) = \frac{\overline{\alpha}+\overline{\beta} - 1}{2-\overline{\beta}-\overline{\alpha}}( 1-\overline{\beta}),\label{eq:ex2}
\end{eqnarray}
to be compared with (\ref{eq:ex1}).
Analogously, in the model considered in section \ref{sec:struct}, for $N_2/N_1 \neq 0$, the map $\overline{\Lambda}^{\infty}$ associates every state $\rho_S$
with the unique fixed point, the vacuum state $|0\rangle \langle 0|$, of the strictly contractive map $\overline{\Lambda}$ in equation (\ref{eq:tag}), i.e.
\begin{equation}
\overline{\Lambda}^{\infty} \rho_S = \left(\begin{array}{cc}
    0&0 \\
    0 & 1
    \end{array} \right),
\end{equation}
and the measure for information trapping reads
\begin{equation}
\mathcal{T}_{\infty}(\overline{\Lambda})= \frac{N_1}{N_1+N_2},
\end{equation}
to be compared with (\ref{eq:gemmer3}). 
In both cases the use of
this alternative measure does not qualitatively change the results,
which justifies to concentrate on idempotence.\\


\begin{thebibliography}{10}

\bibitem{Popescu2006} Popescu A, Short A J and Winter A 2006 {\it{Nat. Phys.}} {\bf{2}} 754
%\bibitem{Goldstein2006} S. Goldstein, J. L. Lebowitz, R. Tumulka, and N. Zangh{\`i}, Phys, Rev. Lett. {\bf{96}}, 050403 (2006)
\bibitem{Reimann2008} Reimann P 2008 {\it{Phys. Rev. Lett.}} {\bf{101}} 190403
\bibitem{Linden2009} Linden N, Popescu A, Short A J and Winter A 2009
  {\it{Phys. Rev.}} E {\bf{79}} 061103
\bibitem{Linden2010} Linden N, Popescu A, Short A J and Winter A 2010
  {\it{New J. Phys.}} {\bf{12}} 055021
\bibitem{Reimann2010} Reimann P 2010 {\it{New J. Phys.}} {\bf{12}} 055027
\bibitem{Lychkovskiy2010} Lychkovskiy O 2010 {\it{Phys. Rev. E}} {\bf{82}} 061103
\bibitem{Short2011} Short A J 2011 {\it{New J. Phys.}} {\bf{13}} 053009
\bibitem{Gogolin2011} Gogolin C, M{\"u}ller M P and Eisert J 2011 {\it{Phys. Rev. Lett}}. {\bf{106}} 040401
\bibitem{Riera2012} Riera A, Gogolin C and Eisert J 2012 {\it{Phys. Rev. Lett.}} {\bf{108}} 080402
\bibitem{Bocchieri1957} Bocchieri P and Loinger A 1957 {\it{Phys. Rev.}} {\bf{107}} 337
\bibitem{Percival1961} Percival I C 1961 {\it{J. Math. Phys.}} {\bf{2}} 235
\bibitem{Fuchs1999} Fuchs C A and van de Graaf J 1999 {\it{IEEE Trans. Inf. Th.}} {\bf 45} 1216
\bibitem{Breuer2007} Breuer H-P and Petruccione F 2007 {\it{The Theory of Open Quantum Systems}}  (Oxford: Oxford University Press)
\bibitem{Kossakowski1972} Kossakowski A 1972 {\it{Rep. Math. Phys.}} {\bf{3}} 247

\bibitem{Ruskai1994a} Ruskai M B 1994 {\it{Rev. Math. Phys.}} {\bf{6}} 1147

\bibitem{Raginsky2002} Raginsky M 2002 {\it{Phys. Rev. A}} {\bf{65}} 032306
\bibitem{Heinosaari2011} Heinosaari T and Ziman M 2011 {\it{The Mathematical Language of Quantum Theory}} (Cambridge: Cambridge University Press)

\bibitem{Breuer2009} Breuer H-P, Laine E-M and Piilo J 2009 {\it{Phys. Rev. Lett.}} {\bf{103}} 210401
\bibitem{Liu2011} Liu B H, Li L, Huang Y-F, Li C-F, Guo G-C, Laine E-M, Breuer H-P and Piilo J 2011 {\it{Nat. Phys.}} {\bf{7}} 93
\bibitem{Laine2010}  Laine E-M, Piilo J and Breuer H-P 2010 {\it{Europhys. Lett.}} \textbf{92} 60010
\bibitem{Rigol2008} Rigol M, Dunjko V and Olshanii M 2008 {\it{Nature}} {\bf{452}} 854
\bibitem{Cramer2008} Cramer M, Dawson C M, Eisert J and Osborne T J 2008 {\it{Phys. Rev. Lett.}} {\bf{100}} 030602
\bibitem{Smirne2010} Smirne A and Vacchini B 2010 {\it{Phys. Rev. A}} {\bf{82}} 022110
\bibitem{Smirne2010b} Smirne A, Breuer H-P, Piilo J and Vacchini B
  2010 {\it{Phys. Rev. A}} {\bf{82}} 062114
\bibitem{Gemmer2004} Gemmer J, Michel M and Mahler G 2004 \emph{Quantum Thermodynamics} (Lecture Notes in Physics {\bf{657}}) (Berlin: Springer)
\bibitem{Gemmer2006} Gemmer J and Michel M 2006 {\it{Eur. Phys. Lett.}} {\bf{73}} 1
\bibitem{Breuer2006} Breuer H-P, Gemmer J and Michel M 2006
  {\it{Phys. Rev. E}} {\bf{73}} 016139
\bibitem{Rivas2010} Rivas A, Huelga S F and Plenio M B 2010 {\it{Phys. Rev. Lett.}} {\bf{105}} 05040
\bibitem{Chruscinski2010} Chru\ifmmode \acute{s}\else \'{s}\fi{}ci\ifmmode \acute{n}\else \'{n}\fi{}ski D, Kossakowski A and Pascazio S 2010 {\it{Phys. Rev. A}} {\bf{81}} 032101
\bibitem{Laine2010b}  Laine E-M, Piilo J and Breuer H-P 2010
  {\it{Phys. Rev. A}} \textbf{81} 062115
\bibitem{Vacchini2011a} Vacchini B, Smirne A, Laine E M, Piilo J, and Breuer H P 2011 {\it{New J. Phys.}} {\bf{13}} 093004
  

\end{thebibliography}
\end{document}